\documentclass[%
 reprint,
superscriptaddress,
 amsmath,amssymb,
]{revtex4-1}

\usepackage{bbold}
\usepackage{mathptmx}
\usepackage{subfig}
\usepackage{psfrag,graphicx}
\usepackage{dcolumn}
\usepackage{amsmath,amssymb}
\usepackage{bm}
\usepackage{color}
\usepackage{latexsym}
\usepackage{epstopdf}
\usepackage{color}
\usepackage[english]{babel}
\usepackage{latexsym}
\usepackage{psfrag,graphicx}
\usepackage{amsfonts}
\usepackage{natbib}
\usepackage{appendix}
\usepackage{aeguill}
\usepackage{ulem}
\usepackage[justification=justified]{caption}

\definecolor{mygrey}{gray}{0.35}
\definecolor{myblue}{rgb}{0.2,0.2,0.8}
\definecolor{myzard}{cmyk}{0,0,0.05,0}
\definecolor{mywhite}{rgb}{1,1,1}
\definecolor{mywhite}{rgb}{1,1,1}
\definecolor{myred}{rgb}{1,0.,0.3}

\usepackage[colorlinks=true,citecolor=myblue,linkcolor=myred]{hyperref}

\def\ba{\begin{align}}
\def\enda{\end{align}}
\def\bi{\begin{itemize}}
\def\ei{\end{itemize}}

\def\be{\begin{equation}}
\def\ee{\end{equation}}
\def\bea{\begin{eqnarray}}
\def\eea{\end{eqnarray}}
\def\bse{\begin{subequations}}
\def\ese{\end{subequations}}

\newcommand{\ket}[1]{|{#1}\rangle}                       
\newcommand{\average}[1]{\langle {#1} \rangle}           

\newcommand{\Ignore}[1]{ }

\def\i{\text{i}}

\begin{document}

\preprint{APS/123-QED}

\title{Level crossings and superradiant quantum phase transition for a two-qutrit quantum Rabi model}


\author{R. Grimaudo}
\address{Department of Physics and Astronomy ``E. Majorana", University of Catania, Via S. Sofia, 64 I-95123 Catania, Italy}

\author{A. S. M. de Castro}
\address{Universidade Estadual de Ponta Grossa, Departamento de F\'{\i}sica, CEP 84030-900, Ponta Grossa, PR, Brazil}

\author{G. Falci}
\address{Department of Physics and Astronomy ``E. Majorana", University of Catania, Via S. Sofia, 64 I-95123 Catania, Italy}
\address{CNR-IMM, UoS Universit\`a, 95123, Catania, Italy}
\address{INFN Sez. Catania, 95123 Catania, Italy}

\author{A. Messina}
\address{ Department of Mathematics and Informatics, University of Palermo, Via Archirafi 34, I-90123 Palermo, Italy}

\author{E. Paladino}
\address{Department of Physics and Astronomy ``E. Majorana", University of Catania, Via S. Sofia, 64 I-95123 Catania, Italy}
\address{CNR-IMM, UoS Universit\`a, 95123, Catania, Italy}
\address{INFN Sez. Catania, 95123 Catania, Italy}

\author{N. V. Vitanov}
\address{Department of Physics, St. Kliment Ohridski University of Sofia, 1164 Sofia, Bulgaria}


\date{\today}

\begin{abstract}
A two-qutrit extension of the quantum Rabi model is studied.
Despite its increased complexity, the model results to be integrable under specific, physically relevant conditions.
This feature allows for the emergence of analytically tractable subdynamics.
In this framework, the ground-state phase diagram can be derived, and the analysis reveals critical phenomena linked to both level crossings and quantum phase transitions.

\end{abstract}



\maketitle

The interest in physical systems with high-dimensional Hilbert spaces is still growing \cite{Adnane24, Zhang20, Jiang24, Adnane24new, Rabbou22, Roy23, Lindon23} and has also extended to the fields of quantum information processing and quantum computation.
In particular, qutrit-based protocols have demonstrated enhanced efficiency in various applications, including quantum gate implementations \cite{FedorovNat12}, quantum cryptography \cite{PasquinucciPRL00, BrussPRL02}, quantum communication \cite{VaziriPRL02}, quantum teleportation \cite{HuPRL20, LuoPRL19}, quantum tomography \cite{LiPRL23}, and quantum key distribution \cite{ErkilicnpjQI23}.
In particular, three-level physical systems offer several advantages over two-level counterparts, such as increased robustness against quantum cloning \cite{BouchardScience17}, reduced experimental demands and resource overheads \cite{LuAQT20}, improved noise resilience \cite{YasirPRA22}, stronger resistance to attacks in quantum key distribution protocols \cite{CerfPRL02}, as well as faster operation and lower error rates in quantum computation \cite{SjoqvistNJP12}.
It is also noteworthy that two-qutrit versions of the Deutsch–Jozsa, Bernstein–Vazirani, and Grover algorithms have recently been proposed \cite{RoyPRApp23}.

Further, over the past decade, few-body systems have attracted significant attention, besides for the pletora of applications listed above, also for the possibility of exhibiting quantum phase transitions (QPTs).
In this case, the spin systems are required to (ultra-)strongly interact with quantized field modes.
This property makes such systems excellent candidates for use as quantum sensors, enabling high-precision measurements of weak physical observables \cite{NiroulaPRL24, DegenRMP17, PatelQIP25}.
Traditionally, QPTs are discussed in the context of many-body quantum systems, where the thermodynamic limit, assuming an infinite number of constituents, is a necessary condition.
However, it has been shown that few-body quantum systems can also undergo QPTs, as demonstrated in the single-qubit \cite{Hwang15, Hwang16} and two-qubit \cite{GdCMSV, GFMPSV} quantum Rabi models (QRMs).
In these cases, new thermodynamic limits must be formulated to establish the appropriate physical conditions for identifying critical points associated with both level crossings and QPTs.

Given the broad range of potential applications of qutrit systems across various areas of physics, it is of significant interest to explore whether open qutrit systems, namely, qutrits coupled to bosonic field modes, can exhibit critical phenomena as well.
In this work, we investigate a two-qutrit extension of the quantum Rabi model, consisting of two interacting qutrits coupled to a single common field mode.
Despite its apparent complexity, by exploiting the symmetry properties of its Hamiltonian, under specific and physically relevant conditions, the model is shown to be integrable and exactly solvable.
This property enables the analytical construction of the phase diagram, revealing the presence of both level crossings and quantum phase transitions.
In the former case, the magnetization serves as the order parameter, while in the latter, the mean photon number and the entanglement between the two qutrits act as probes of the critical behavior.
This extends the class of few-body quantum systems known to exhibit quantum phase transitions.
Further, it supports the notion that multiple definitions of suitable thermodynamic limits may apply, and broadens the potential use of qutrit systems as quantum sensors.

\textit{Two-Qutrit QRM.}
Let us consider the following model of two interacting qutrits coupled to the same bosonic mode
\begin{equation} \label{Hamiltonian}
\begin{aligned}
H &=
\Omega_{1}\hat{\Sigma}_{1}^{z}+\Omega_{2}\hat{\Sigma}_{2}^{z}
+\gamma_{x}\hat{\Sigma}_{1}^{x}\hat{\Sigma}_{2}^{x}
+\gamma_{y}\hat{\Sigma}_{1}^{y}\hat{\Sigma}_{2}^{y}
+\gamma_{z}\hat{\Sigma}_{1}^{z}\hat{\Sigma}_{2}^{z} + \\
&+ \omega \hat{a}^\dagger \hat{a} + 
\sum_{k=1}^2 \lambda_{k} \left( \hat{a}^\dagger + \hat{a} \right) \hat{\Sigma}_k^z,
\end{aligned}
\end{equation}
where $\Omega_k$ ($k=1,2$) are the characteristic frequencies of the two qutrits, and $\omega$ is the frequency related to the field mode.
$\lambda_{k}$ are the coupling constants characterizing the qutrit-mode interaction, and $\gamma$s are the different energy contributions resulting from the coupling between the two three-level systems.
$\hat{\Sigma}_i^k$ ($k=x,y,z$) are the spin-1 angular momentum operators, while $\hat{a}^\dagger$ and $\hat{a}$ are the standard boson field operators satisfying the commutation relation $[ \hat{a},\hat{a}^\dagger]=1$.

Recent works have shown that several quantum platforms can support controlled interactions between qutrits—and between qutrits and quantized field modes.
In superconducting circuits, both theoretical proposals and experiments (e.g., using transmon qutrits) demonstrate that the third energy level can mediate resonant or dispersive interactions, enabling cavity-assisted state transfer, GHZ-state generation, and high-fidelity two-qutrit entangling gates \cite{Yang12, Liu20, Goss22}.
Similar cavity-mediated schemes have been proposed for transferring arbitrary multilevel states in circuit-QED architectures \cite{Liu17new}.
Multilevel NV centers coupled to optical resonators offer another route, where Raman processes and generalized Jaynes–Cummings interactions enable spin–photon coupling in three-level configurations \cite{Li11}.
More recently, dissipative strategies have been proposed to stabilize entangled qutrit subspaces through shared dispersive coupling to a resonator \cite{Wang23}.
In trapped-ion systems, qutrit–qutrit interactions are typically mediated by collective motional modes, allowing precise control of multilevel dynamics and native qudit gates \cite{Hrmo23}.
Together, these results highlight that multiple platforms now support viable mechanisms for implementing qutrit-qutrit and qutrit-mode interactions.

Based on previous results \cite{GMIV, GVM1}, the Hamiltonian in Eq. \eqref{Hamiltonian} is shown to present the following constant of motion:
\begin{equation}\label{Cos Form of K}
\hat{K} = \cos(\pi\hat{\Sigma}_{\rm tot}^z),
\end{equation}
where $\hat{\Sigma}_{\rm tot}^z = \hat{\Sigma}_{1}^{z} + \hat{\Sigma}_{2}^{z}$.
As a consequence, two dynamically invariant Hilbert subspaces, related to the two eigenvalues ($\pm 1$) of $\hat{K}$, exist: $\mathcal{H}_-$, spanned by $\{\ket{10},~\ket{01},~\ket{0-1},~\ket{-10}\} \otimes \{ \ket{n} \}_n$; and $\mathcal{H}_+$, spanned by $\{\ket{11},~\ket{1-1},~\ket{00},~\ket{-11},~\ket{-1-1}\} \otimes \{ \ket{n} \}_n$.
$\{\ket{1},~\ket{0},~\ket{-1}\}$ are the eigenstates of $\hat{\Sigma}^z$, while $\ket{n}$ are the Fock states defined by $\hat{a}^\dagger \hat{a} \ket{n} = n \ket{n}$, with $n \in \mathrm{N}$.

The projection of the Hamiltonian $H$ into the subspace $\mathcal{H}_-$ can be written as
\begin{equation}\label{4x4 block as two spin-1/2}
H_0 = H_{a} + H_{b} + H_o,
\end{equation}
with
\begin{equation}\label{H1 and H2}
\begin{aligned}
H_{a} &= \Omega_+\hat{\sigma}_{a}^{z}+
\gamma_-\hat{\sigma}_{a}^{x} + 
\lambda_+ \left( \hat{a}^\dagger + \hat{a} \right) \hat{\sigma}_a^z,
\\
H_{b} &= \Omega_-\hat{\sigma}_{b}^{z}+
\gamma_+\hat{\sigma}_{b}^{x} + 
\lambda_- \left( \hat{a}^\dagger + \hat{a} \right) \hat{\sigma}_b^z,
\\
H_o &= \omega \hat{a}^\dagger \hat{a},
\end{aligned}
\end{equation}
where $\hat{\sigma}^k$ ($k=x,y,z$) are spin-1/2 Pauli matrices and we set $\Omega_\pm=(\Omega_1\pm\Omega_2)/2$, $\gamma_\pm=\gamma_x\pm\gamma_y$ and $\lambda_\pm=(\lambda_{1} \pm \lambda_{2})/2$.
The appearance of effective two-level operators results from the following mapping between the four two-qutrit states involved in the subspace $\mathcal{H}_-$ and the standard four two-qubit states ($\ket{\pm}$ being the two eigenstates of $\hat{\sigma}^z$):
\begin{equation}\label{Mapping}
\begin{aligned}
\ket{10} & \hspace{0,25cm} \leftrightarrow \hspace{0,25cm} \ket{++},
\qquad
\hspace{0,275cm} \ket{01} \hspace{0,25cm} \leftrightarrow \hspace{0,25cm} \ket{+-},\\
\ket{0-1} & \hspace{0,25cm} \leftrightarrow \hspace{0,25cm} \ket{-+},
\qquad
\ket{-10} \hspace{0,25cm} \leftrightarrow \hspace{0,25cm} \ket{--}.
\end{aligned}
\end{equation}
$H_0$ represents two non-directly interacting two-level systems coupled to a common field mode.
We stress that the boson operators in $H_0$ are formally different from the ones in Eq. \eqref{Hamiltonian}, since in Eq. \eqref{H1 and H2} they must be intended multiplied by the operator projecting to the subspace $\mathcal{H}_-$, and then they do not act on the orthogonal subspace $\mathcal{H}_+$.

The effective model of the two qubits indirectly interacting via the field mode can be approached and solved on the basis of the methods and the results existing in literature \cite{Peng14, Wang14, Duan15, Mao15, Dong16, Sun20}.
However, such solutions cannot be written in a closed form.
In the following, instead, we are interested in identifying specific and physically meaningful conditions, making the exact solutions expressible in a closed analytical form.

By considering the following relations between the Hamiltonian parameters 
\begin{equation} \label{Conditions1}
  \begin{aligned}
    \Omega_1 = \Omega_2 = \Omega
    \qquad
    \gamma_x = \gamma_y = \gamma
    \qquad
    \lambda_1 = \lambda_2 = \lambda,
  \end{aligned}
\end{equation}
the mathematical expressions of $H_a$ and $H_b$ considerably simplify:
\begin{equation}\label{H1 and H2 simplified}
\begin{aligned}
H_{a} &= \Omega \hat{\sigma}_{a}^{z}+
\lambda \left( \hat{a}^\dagger + \hat{a} \right) \hat{\sigma}_a^z,
\\
H_{b} & = \gamma \hat{\sigma}_{b}^{x},
\qquad
H_o = \omega \hat{a}^\dagger \hat{a}.
\end{aligned}
\end{equation}
It is worth noticing that the conditions in Eq. \eqref{Conditions1} strengthen the symmetry of the Hamiltonian model \eqref{Hamiltonian} since it now commutes also with $\hat{\Sigma}_{\text{tot}}^z$.
This circumstance induces a further decomposition of the Hilbert space: namely, $\mathcal{H}_-$ ($\mathcal{H}_+$) splits into two (three) invariant subspaces.


In this instance, the analytical solutions of the effective two-qubit-mode model can be easily derived (this time in a closed form) and, basing on the mapping in Eq. \eqref{Mapping}, they can be written in terms of the two-qutrit-mode states as follows
\begin{equation} \label{Eigenvectors H-}
  \begin{aligned}
    \ket{\Psi_n^{+\pm}} &= {\ket{10} \pm \ket{01} \over \sqrt{2}} \otimes \ket{-\alpha, n},
    \\
    \ket{\Psi_n^{-\pm}} &= {\ket{0-1} \pm \ket{-10} \over \sqrt{2}} \otimes \ket{\alpha, n},
  \end{aligned}
\end{equation}
with corresponding eigenvalues
\begin{equation} \label{Eigenvalues H-}
  \begin{aligned}
    \epsilon_n^{+\pm} &= \Omega \pm \gamma + (n - \alpha^2) \omega,
    \\
    \epsilon_n^{-\pm} &= -\Omega \pm \gamma + (n - \alpha^2) \omega,
  \end{aligned}
\end{equation}
where $\alpha=\lambda/\omega$, and $\ket{\pm \alpha, n} = D(\pm \alpha) \ket{n}$, with $D(\alpha) = \exp\{ \alpha(\hat{a}^\dagger - \hat{a}) \}$ being the bosonic displacement operator.
We stress that the two above pairs of eigenstates and eigenvalues belong to the two different invariant subspaces related to the two eigenvalues $+1$ and $-1$ of  $\hat{\Sigma}_{\text{tot}}^z$, respectively.

If we consider now the subspace related to the eigenvalue $0$ of $\hat{\Sigma}_{\text{tot}}^z$ (spanned by $\{ \ket{1-1}, \ket{00}, \ket{-11} \} \otimes \{ \ket{n} \}_n$) the effective Hamiltonian can be written as
\begin{equation}\label{H3}
H_3=
\sqrt{2} \gamma \hat{\Sigma}^{x} +
\omega \hat{a}^\dagger \hat{a}.
\end{equation}
The above expression has been obtained by considering $\gamma_z=0$, besides the conditions in eq. \eqref{Conditions1}.
We emphasize that the choice $\gamma_z=0$ does not alter the sub-dynamics in the subspace $\mathcal{H}_-$ since $H_a$ and $H_b$ in Eq. \eqref{H1 and H2} do not depend on $\gamma_z$.
The eigensolutions of $H_3$ can be trivially written as
\begin{equation}
  \begin{aligned}
    \ket{\Phi_n^\pm} &= {\ket{1-1} \pm\sqrt{2} \ket{00} + \ket{-11} \over 2} \otimes \ket{n},
    \\
    \ket{\Phi_n^0} &= {\ket{1-1} - \ket{-11} \over \sqrt{2}} \otimes \ket{n},
  \end{aligned}    
\end{equation}
with corresponding eigenvalues
\begin{equation}
  \begin{aligned}
    \eta_n^\pm = \pm \sqrt{2} \gamma + n \omega
    \qquad
    \eta_n^0 = n \omega.
  \end{aligned}    
\end{equation}
The last eigenstates belonging to two eigenspaces related to the eigenvalues $+2$ and $-2$ of $\hat{\Sigma}_{\text{tot}}^z$ are, respectively,
\begin{equation}
\begin{aligned}
    \ket{\Theta_n^+} &= \ket{11} \otimes \ket{-2\alpha, n} \\
    \ket{\Theta_n^-} &= \ket{-1-1} \otimes \ket{2\alpha, n},
\end{aligned}
\end{equation}
with eigenenergies
\begin{equation}
\begin{aligned}
    e_n^+ &= 2\Omega + (n - 4\alpha^2) \omega \\
    e_n^- &= -2\Omega + (n - 4\alpha^2) \omega.
\end{aligned}
\end{equation}

\begin{figure}[] 
\begin{center}
{\includegraphics[width=0.45\textwidth]{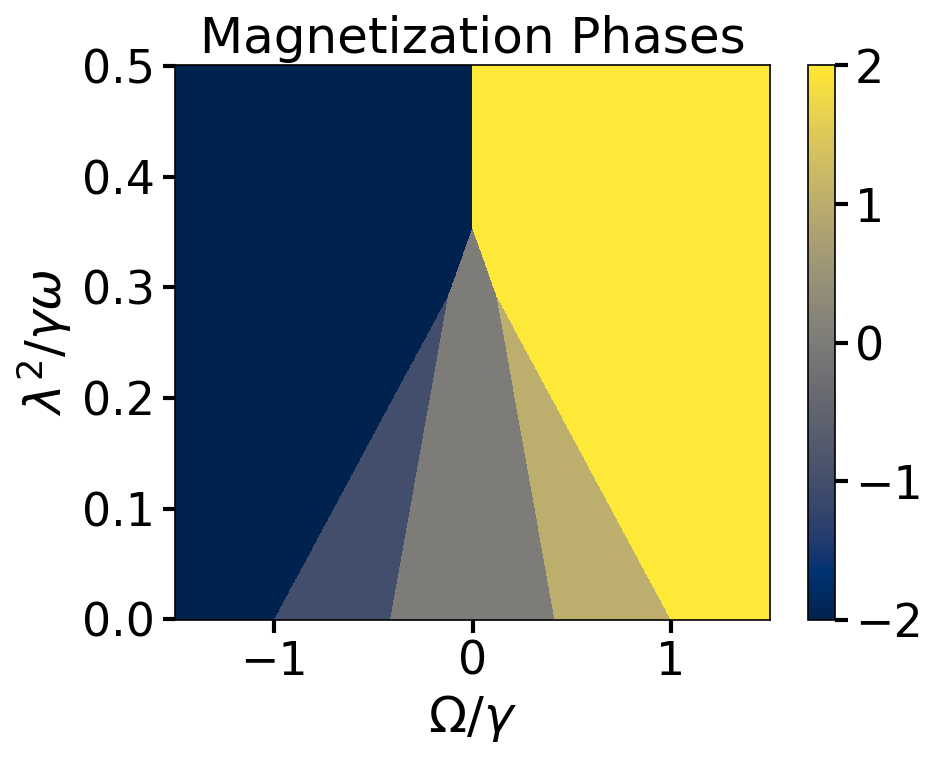}}
\captionsetup{justification=raggedright,format=plain,skip=4pt}%
\caption{Magnetization phase diagram of the two-qutrit QRM \eqref{Hamiltonian} under the conditions \eqref{Conditions1}.
The different colors related to different values of the total two-qutrit magnetization correspond to different ground states of the two-qutrit-mode system.}
\label{fig: qrmII_energies}
\end{center}
\end{figure}

\textit{Level Crossings.}
The ground state of the system depends substantially on the region of the parametric space defined by the two dimensionless quantities $\Omega/\gamma$ and $\alpha^2\omega/\gamma=\lambda^2/\gamma\omega$, under the conditions \eqref{Conditions1}.
In Fig. \ref{fig: qrmII_energies} five phases can be indeed identified related to five different states of the two-qutrit-mode system and belonging to different eigenspaces of the constant of motion $\hat{\Sigma}_{\text{tot}}^z$.
Each of these phases has then a specific fixed value of the total magnetization, as shown in Fig. \ref{fig: qrmII_energies}.
The edge lines separating the different phases consist in level-crossing points, that is, critical points where two or more eigenstates are degenerate.
It is possible to identify indeed three tri-critical points, that is, points where three eigenstates have the same energy, namely at $(\Omega/\gamma, ~ \lambda^2/\gamma\omega) = (0, ~ 0.34)$ and $(\Omega/\gamma, ~\lambda^2/\gamma\omega) \approx (\pm 0.4, ~ 0.28)$.

We note that for $\lambda^2/\gamma\omega \lesssim 0.28$, by changing $\Omega/\gamma$ from 0 to $>1$ the system explores five phases related to the five ground states (in order) $\ket{\Theta_0^-}$, $\ket{\Psi_0^-}$, $\ket{\Phi_0^-}$, $\ket{\Psi_0^+}$, $\ket{\Theta_0^+}$.
For $\lambda^2/\gamma\omega \gtrsim 0.34$, instead, only two phases exist, with just one consequent crossing from $\ket{\Theta_0^-}$ to $\ket{\Theta_0^+}$.
In the middle region, that is, for $0.34 \gtrsim \lambda^2/\gamma\omega \gtrsim 0.28$, the phase related to $\ket{\Phi_0^-}$ appears between the last two.

By fixing $\Omega/\gamma$ and changing $\lambda^2/\gamma\omega$, instead, at most three phases can be crossed.
We point out that, besides the magnetization, the different phases can be distinguished by other two meaningful physical quantities: the mean photon number ($N$) and the level of entanglement exhibited by the two qutrits.
The latter can be quantified by calculating the negativity ($\mathcal{N}$) of the ground states \cite{Vidal02, Lee03}.

The five phases related to the eigenvalues of the magnetizations $M=\pm2, ~ \pm 1, ~ 0$ are indeed characterized by a mean photon number $N=4\alpha^2, ~ \alpha^2, ~ 0$, and a level of entanglement equal to $\mathcal{N} = 0, ~ 1/2, ~ (1+2\sqrt{2})/4$, respectively.
It is worth stressing that $\ket{\Psi_0^{+-}}$ and $\ket{\Psi_0^{--}}$, the ground states related to $M=\pm 1$, exhibit the maximum level of entanglement reachable in those subspaces (namely, 1/2) \cite{GMIV, GVM1}.
The maximum value of negativity is however reached in the phase with $\ket{\Phi_0^{-}}$ as ground state, characterized by $M=0$.

Finally, we point out that such a phase diagram considers qutrit-mode couplings which can be either weak ($\lambda/\omega < 0.1$) or strong ($\lambda/\omega > 0.1$) thanks to the fact that the validity of our approach and the exact analytical solutions derived above are independent of the qutrit-mode coupling strength.
\\\\

\textit{QPT.}
Let us consider a new scenario identified by the following slightly different physical conditions: 
\begin{equation} \label{Conditions}
  \begin{aligned}
    \Omega_1 = \Omega_2 = \Omega
    \quad
    \gamma_x = \gamma_y = \gamma/2
    \quad
    \lambda_1 = - \lambda_2 = \lambda/2.
  \end{aligned}
\end{equation}
We stress that we are considering now an opposite coupling between the two qutrits with the mode.
This choice is not a mere mathematical curiosity, but, as we can see in the following, the system exhibits a relevant different physics stemming from the different symmetry characterizing the coupling between the two qutrits and the field.

Under the new conditions \eqref{Conditions}, the Hilbert subspace $\mathcal{H}_-$ is described by the effective two-qubit-mode Hamiltonian
\begin{equation}\label{H1 and H2 simplified 2}
\begin{aligned}
H'_{\sigma_a^z} = \sigma_a^z \Omega + {\gamma \over 2} \hat{\sigma}_{b}^{x} + 
{\lambda \over 2} \left( \hat{a}^\dagger + \hat{a} \right) \hat{\sigma}_b^z +
\omega \hat{a}^\dagger \hat{a}.
\end{aligned}
\end{equation}
In the above expression $\sigma_a^z$ is a constant of motion and two equivalent Hamiltonians are then present ($\sigma_a^z = \pm 1$).
The Hamiltonian $H_3$ becomes instead
\begin{equation}\label{H3 new}
H_3' = {\gamma \over \sqrt{2}} \hat{\Sigma}^{x} + \omega \hat{a}^\dagger \hat{a} + 
{\lambda} \left( \hat{a}^\dagger + \hat{a} \right) \hat{\Sigma}^z.
\end{equation}

We note that $H_{\sigma_a^z}'$ has the form of a QRM.
It is well known that the latter exhibits a second-order QPT when the ratio of the qubit frequency to that of the mode goes to infinity \cite{Hwang15}.
Such a limit (not related to the size of the system) plays the role that the thermodynamic limit plays in many-body spin systems \cite{Bakemeier12, Ashhab13}.
Differently from many-body systems, other thermodynamic-like limits can be defined in order to ensure the classical oscillator limit in few-body quantum systems \cite{GFMPSV}.

Analogously to the two-qubit QRM \cite{GFMPSV}, in the case of the tripartite system analysed in this work, we can recover the thermodynamic limit by considering the infinite ratio of the qutrit-qutrit coupling to the mode frequency, namely
\begin{equation} \label{limits}
\gamma/\omega \rightarrow \infty.
\end{equation}
Such a limit, although being profoundly physically different from the one introduced for the single-qubit QRM \cite{Bakemeier12, Ashhab13, Hwang15}, can be seen mathematically equivalent to the latter.
This circumstance stems from the fact that in the effective qubit model $H_{\sigma_a^z}'$, describing the two-qutrit-mode dynamics in the invariant subspace $\mathcal{H}_-$, the role of the frequency of the fictitious qubit is played exactly by the qutrit-qutrit interaction parameter $\gamma$. 
Being interested in the ground state of the two-qutrit-mode system and its properties, from now on we consider only the low-energy Hamiltonian $H'_{\sigma_a^z=-1} \equiv H_{-}'$, supposing $\Omega>0$. 

\begin{figure}[htp]
\begin{center}
\subfloat[][]{\includegraphics[width=0.22\textwidth]{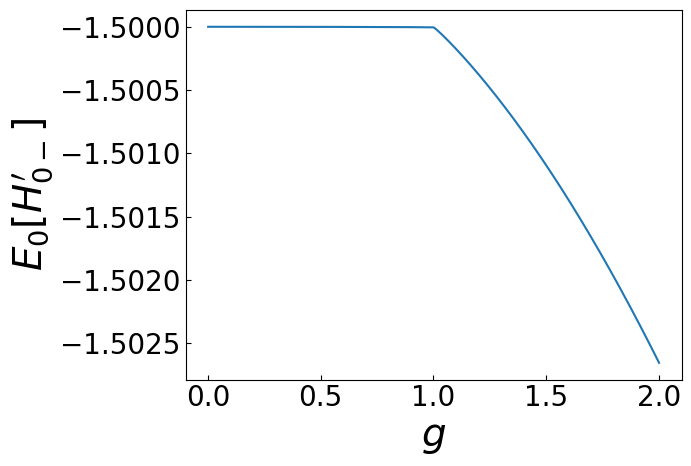}\label{fig: GS H4}}
\qquad
\subfloat[][]{\includegraphics[width=0.22\textwidth]{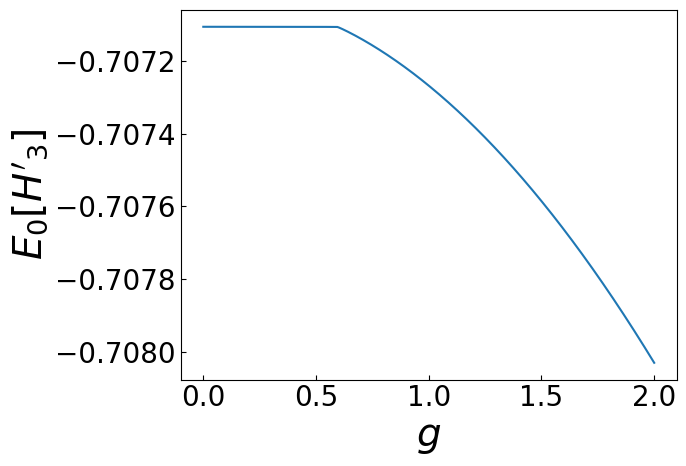}\label{fig: GS H3}}
\captionsetup{justification=raggedright,format=plain,skip=4pt}%
\caption{Numerical solution, under conditions in Eq. \eqref{Conditions}, of the ground state energy of (a) $H_{0-}'$ and (b) $H_3'$, for $\omega/\gamma = 10^{-6}$ and $\Omega/\gamma=1$.}
\label{fig: GS comparison}
\end{center}
\end{figure}

Of course, we have to pay attention on the possibility that the ground state of the system belongs to the subspace governed by $H'_3$.
That is, $H'_3$ may present a lower energy state with respect to the lowest energy state of $H_{-}'$.
Nevertheless, thanks to the presence of the parameter $\Omega$ in $H'_3$, the latter can be chosen in order that the lowest energy state belongs to the subspace ruled by $H_{-}'$, as shown in Fig. \ref{fig: GS comparison}, when it has been set $\Omega/\gamma = 1$ and $\omega/\gamma = 10^{-6}$.
The abrupt change of the dependence of the ground energy on $g$ at $g=1$ is a clear signature of the presence of a QPT.
In the following, the presence of such a QPT is analytically derived.

On the basis of the results obtained for the single-qubit \cite{Hwang15} and the two-qubit QRM \cite{GFMPSV}, under the physical condition expressed in Eq. \eqref{limits}, the Hamiltonian $H'_{-}$ acquires the form
\begin{equation} \label{Ham np}
\begin{aligned}
{H}_{-}^{np} = \pm \Omega +
\omega ~ a^\dagger a -
{\omega g^2 \over 4} (a^\dagger + a)^2 - {\gamma \over 2},
\end{aligned}
\end{equation}
for $g<1$, and
\begin{equation} \label{Ham sp}
\begin{aligned}
{H}_{-}^{sp} = \pm \Omega + 
\omega ~ a^\dagger a -
{\omega \over 4 g^4} (a^\dagger + a)^2 -
\gamma ~ {g^2 + g^{-2} \over 4},
\end{aligned}
\end{equation}
for $g>1$, where the parameter $g$ has been defined as $g=2\lambda/\sqrt{\omega\gamma}$.

The superscripts in the two Hamiltonians $H_{-}^{np}$ and $H_{-}^{sp}$, being the exact form of the Hamiltonian $H'_{-}$ for the low-energy spectrum \cite{Hwang15}, are related to two different phases: the normal phase ($np$, $g<1$) and the superradiant one ($sp$, $g>1$), related to the normalized mean photon number of the related ground states: vanishing in the normal phase and non-zero in the superradiant one.
Such a criticality presented by $H'_{-}$ for $g=1$ means that the two-qutrit QRM, as the single- \cite{Hwang15} and the two-qubit \cite{GFMPSV} QRM, presents a superradiant QPT.  

We remark that, besides $\gamma/\omega \rightarrow \infty$, the limit $\lambda/\omega \rightarrow \infty$ must be contextually considered in order to have a non-vanishing control parameter $g$, driving the system towards the QPT; otherwise, $g$ would vanish, making it impossible the occurrence of the QPT \cite{Hwang15}.
Thus, thanks to the dynamical decomposition and the reinterpretation of the two subdynamics in terms of easier fictitious systems, we have transparently highlighted the occurrence of a QPT in our tripartite two-qutrit QRM.

The exact expressions of the lowest-energy states of $H_{-}^{np}$ and $H_{-}^{sp}$ (that is, of $H'_{-}$ in the two distinct regimes) and the related eigenenergies can be analytically derived \cite{Hwang15, GFMPSV}.
Basing then on the mapping in Eq. \eqref{Mapping}, we can write the ground states in terms of the states of the two-qutrit-mode system as follows:
\begin{equation} \label{GS np}
\begin{aligned}
E_{0-}^{np} &= \omega {\sqrt{1-g^2} - 1 \over 2} - {\gamma \over 2} - \Omega,
\\
\ket{\Psi_{0-}^{np}} &= \mathcal{S}[r_{np}(g)] \ket{0} \otimes {\ket{0-1}-\ket{-10} \over \sqrt{2}},
\end{aligned}
\end{equation}
for $g<1$, and
\begin{equation} \label{GS sp}
\begin{aligned}
E_{0-}^{sp} =& ~ \omega {\sqrt{1-g^{-4}} - 1 \over 2} - \gamma{g^2 + g^{-2} \over 4} - \Omega,
\\
\ket{\Psi_{0-}^{sp}} =& ~ \mathcal{S}[r_{sp}(g)] \ket{0} \otimes \\
 \Bigg[& \left( {\sqrt{1+g^{-2}} \mp \sqrt{1-g^{-2}} \over 2} \right) \ket{0-1} - \\
& \left( {\sqrt{1+g^{-2}} \pm \sqrt{1-g^{-2}} \over 2} \right) \ket{-10} \Bigg],
\end{aligned}
\end{equation}
when $g>1$, with $r_{np} = - \ln(1-g^2)/4$, $r_{sp} = - \ln(1-g^{-4})/4$, and $\mathcal{S}(x) = \exp\{ (x/2) (a^{\dagger2} - a^2) \}$.
Note that the ground state $\ket{\Psi_{0-}^{sp}}$ of $H^{sp}_{-}$ in Eq. \eqref{GS sp} is two-fold degenerate.

If we introduce the rescaled energy $\widetilde{E}=E \cdot \omega/\gamma$, in the limit $\gamma/\omega \rightarrow \infty$, we get
\begin{equation} \label{rescaled energies}
\begin{aligned}
\widetilde{E}_0^- &=
\begin{cases}
\widetilde{E}_{0-}^{np} = - \omega, \quad g<1
\\\\
\widetilde{E}_{0-}^{sp} = - \omega ~ ({g^2 + g^{-2}) / 2}, \quad g>1.
\end{cases}
\end{aligned}
\end{equation}

\Ignore{
\begin{figure}[] 
\begin{center}
\includegraphics[width=0.45\textwidth]{sr_IIqpt_2qqrm_1.pdf}
\captionsetup{justification=raggedright,format=plain,skip=4pt}%
\caption{Dependence of the first two lowest eigenenergies $\widetilde{E}_0^+/\omega$ and $\widetilde{E}_0^-/\omega$ on the control parameter $g=g_1=g_2$, for counter-biased qubits ($\epsilon_1=-\epsilon_2)$ and equal spin-mode couplings ($\lambda_1=\lambda_2$).
The black curve in the superradiant phase ($g>1$) corresponds to two degenerate eigenstates (see Eq. \eqref{GS sp}).
(b) Dependence of the rescaled mean photon number on $g$, for the ground state of the two-qubit QRM, with $\epsilon / \gamma \rightarrow 0$, $\omega / \gamma \rightarrow 0$, and $\omega / \lambda \rightarrow 0$.}
\label{fig: qrmII_multiplot_1}
\end{center}
\end{figure}

\begin{figure}[] 
\begin{center}
\includegraphics[width=0.45\textwidth]{sr_IIqpt_2qqrm_2.pdf}
\captionsetup{justification=raggedright,format=plain,skip=4pt}%
\caption{
Dependence of the two-qubit concurrence (a), and the two-qubit magnetization (b) on $g$, for the ground state of the two-qubit QRM, with $\epsilon / \gamma \rightarrow 0$, $\omega / \gamma \rightarrow 0$, and $\omega / \lambda \rightarrow 0$.
The two curves (black and blue) for $g>1$ (superradiant phase) are related to the two states in Eq. \eqref{GS sp}
}
\label{fig: qrmII_multiplot_2}
\end{center}
\end{figure}
}

It is worth pointing out that, given the factorization of the ground states in qubit and bosonic parts (for $g<1$ and for $g>1$), the mean photon number $N=\average{\Psi_0|a^\dagger a|\Psi_0}$, the two-qutrit magnetization $M=\average{\Psi_0|(\Sigma_1^z+\Sigma_2^z)/2|\Psi_0}$ and the negativity $\mathcal{N}$ \cite{Wootters98} can be easily obtained (here $\ket{\Psi_0}$ indicates the generic ground state, independently of the phase, either normal or superradiant).
In the following these three quantities both for the normal ($g<1$) and the superradiant ($g>1$) phases are given,
\begin{equation} \label{physical quantities}
\begin{aligned}
\widetilde{N}_{np} = 0, &\qquad \widetilde{N}_{sp} = {g^2 - g^{-2} \over 4},\\
\mathcal{N}_{np} = {1 \over 2}, &\qquad \mathcal{N}_{sp} = {g^{-4} \over 2}, \\
M_{np} = 0, &\qquad M_{sp} = \pm \sqrt{1 - g^{-2}},
\end{aligned}
\end{equation}
where $\widetilde{N} = N \cdot (\omega/\gamma)$ is the rescaled mean photon number.
The rescaling is necessary in order to highlight its difference in the two phases.
In fact, for $g>1$ the mean photon number is infinite, as usual for the superradiant phase. 

\begin{figure}[htp]
\begin{center}
\subfloat[][]{\includegraphics[width=0.22\textwidth]{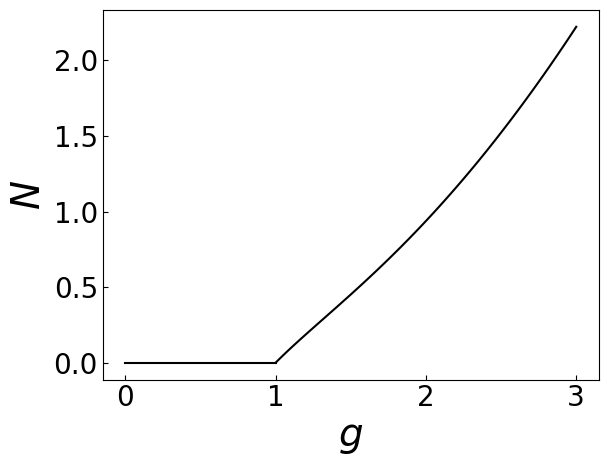}\label{fig: mpn}}
\qquad
\subfloat[][]{\includegraphics[width=0.22\textwidth]{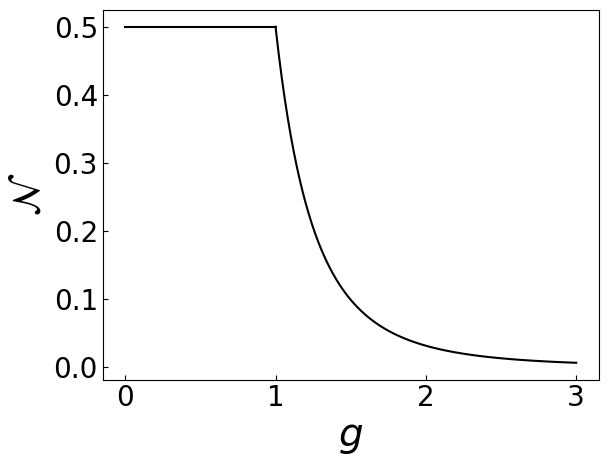}\label{fig: neg}}
\captionsetup{justification=raggedright,format=plain,skip=4pt}%
\caption{Dependence on $g$ of (a) the rescaled mean photon number and (b) the negativity (level of entanglement between the two qutrits) for the ground state of the system under the conditions in Eq. \eqref{Conditions}. The abrupt change of the dependence on $g$ is related to the QPT, that is the change of the ground state of the system.}
\label{fig: mpn and neg}
\end{center}
\end{figure}

The drastic change of the dependence of the mean photon number on $g$ (see Fig. \ref{fig: mpn}) is at the origin of the superradiant nature of the QPT.
Nevertheless, the two phases are characterized also by a different level of entanglement exhibited by the two qutrits.
Precisely, the normal phase is characterized by the maximum level of entanglement admissible in that subspace ($\mathcal{N}_{np}=1/2$) \cite{GMIV}, while the superradiant phase exhibits a negativity which decreases as $g$ increases (see Fig. \ref{fig: neg}).
Therefore, like in other spin systems \cite{Osterloh02, Oliveira06, Gu07, Chen16, Yuste18}, besides the mean photon number, also the level of entanglement between the two qutrits can serve as order parameter for the onset of the superradiant QPT.
Both quantities $\partial N / \partial g$ and $\partial \mathcal{N} / \partial g$ present indeed a discontinuity at the critical point $g=1$. 
Other signatures confirming the presence and characterizing the class of such a second order QPT are the scaling behaviours of the energy gap between the ground and the first excited state, as well as of the variance of both position and momentum quadrature of the cavity mode \cite{Hwang15}.

\textit{Conclusions.}
We have demonstrated the dynamical richness of the generalized two-qutrit version of the QRM.
Owing to its integrability, the analysis of this model can be decomposed into simpler subproblems, some of which become exactly solvable under specific and physically meaningful conditions on the Hamiltonian parameters.
Consequently, more complex generalized QRMs can, perhaps counterintuitively, prove to be more mathematically tractable than the standard single-qubit QRM.
This property enables the analytical identification of notable physical features, such as the emergence of critical behaviors associated with both level crossings and quantum phase transitions in the ground states.
As a result, these systems represent a valuable platform for investigating fundamental theoretical aspects and exploring potential experimental implementations in quantum technologies, such as quantum sensing.

\textit{Acknowledgments.}
This work was supported by the PNRR MUR Project
No. PE0000023-NQSTI.
G.F. and E.P. acknowledge the QuantERA grant SiUCs (Grant No. 731473) and the grant Pia.Ce.Ri.-UNICT, project Q-ICT.
A.S.M.C. acknowledges support from Fundação Araucária (Project No. 305).

\bibliography{biblio_qrm}

\end{document}